# Thermomechanical Assessment of Breast Tumor Subjected to Focused Ultrasound and Interstitial Laser Heating

Abhijit Paul, Anup Paul*

*Abstract*— During laser and ultrasound thermotherapy it is always desirable to have a precise necrosis of deeply seated tumor preserving the adjoining healthy tissue with minimum thermally induced nociceptive pain sensation to the patient. The aim of the present study is to determine the effects of nanoparticle mixed tissues under pulsed and continuous heating during high intensity focused ultrasound (HIFU) and laser interstitial thermal therapy (LITT). The present problem incorporating the tissue thermal relaxation times ($\tau_q, \tau_T$) was solved in a 3-dimensional multilayered vasculature breast tumor model signifying the complex inhomogeneous tissue structure. The coupled Radiative transfer, Helmontz, momentum, dual phase lag (DPL) and equilibrium equations for optic, acoustic, fluid, temperature and mechanical fields respectively were solved simultaneously using COMSOL Multiphysics (Bangalore, India) software. An in-vitro study on agar based tissue phantom was also performed to validate the present numerical results of focused ultrasound heating. The thermal relaxation times of tissue causes significant changes of thermal and damage history under focused ultrasound heating compared to unfocused laser heating. With limited rise in tissue temperature, the pulsed mode of heating for longer period with lower duty cycle (16.6%) shows a target specific necrotic damage with reduced nociceptive pain in contrast to continuous mode of heating. Further, the presence of nanoparticles and multilevel artery and vein affect both the thermal and mechanical response under external heating. Thus, the present findings could help to understand the role of different external heating modes and sources on tumor necrosis during clinical practice of thermo-therapy.

I. INTRODUCTION

IN recent days breast cancer is considered as common in women and its chances of lethality comes second after lung cancer [1]. Therefore non-invasive therapeutic methods viz. laser assisted thermo-therapy [2-7], focused ultrasound ablation [8-12], etc. have gained attention for treatment of deeply-seated tumors which are very challenging to remove completely through surgery. As a minimally invasive therapeutic method, laser interstitial thermal therapy (LITT) is widely accepted for clinical treatment of liver, breast, prostate and brain cancer [2] guided by computational prediction [3]. During laser assisted thermal therapy, overheating of target region is main concern. However, the overheating can be eliminated to some extent using pulsed laser [4-7]. Through in-vitro and numerical study authors [4, 5] have revealed that by pulsed laser higher tissue temperature can be achieved comparing continuous wave laser. Other studies [6, 7] show that the laser pulse duration can affect the elevation of tissue therapeutic temperature. Simanovski et al [6] proposed that the threshold temperature of cell viability decreases with laser pulse duration. In a pioneering work by Jain et al [7] witnessed that the post therapy lesion size is a function of laser power and pulse width.

Focused ultrasound is the sole non-invasive, conformal and nonradiative [8-12] technique to ablate local tumor placed deeply inside the body avoiding severe damage to the skin and adjacent healthy tissue. As a recently developing thermal therapy, high intensity focused ultrasound (HIFU) has gained a wider attention [8-11] in clinical as well as research practices. During HIFU the tumor damage mechanism mainly consists of cavitation, thermal and mechanical effects [8] and concise necrosis [12] of target region can be achieved in a more controlled form. The real-time temperature monitoring and evaluation of treatment efficacy is a key challenge for HIFU therapy [9, 10, 13]. Kim et al. [13] had proposed a novel integrated system of photoacoustic thermometry to monitor HIFU treatment. HIFU can be applied in both continuous and pulsed train form. In continuous mode, it can increase the tissue temperature beyond 70 ℃ very rapidly within seconds and could be applied clinically to ablate carcinomas tumors [14]. However during continuous treatment, it is difficult to control such a rapid temperature elevation in real time [15] which can result unwanted healthy tissue damage. Nevertheless HIFU in pulsed mode with lower duty cycle (≤10%) can provide better control [16] over rapid temporal tissue temperature rise.

In spite of higher clinical acceptance of HIFU, problem still persists such as overheating of surrounding normal tissue and skin damage [11]. To avoid such problems, the intensity of acoustic energy for complete ablation can be reduced by introducing energy absorbing materials. Hence, different micron to nano level bubbles [17] and different type of

Abhijit Paul is with the Department of Mechanical Engineering, NIT Arunachal Pradesh, 791112, India,(email:abhijitpaul1501@gmail.com).
*Anup Paul is with the Mechanical Engineering Department, NIT Arunachal Pradesh, 791112, India (e-mail: catchapu@nrim.go.jp).

nanoparticles (Nps) [11] are infused into the target site. The NPs, due to their nano size can easily accumulate at the targeted cells [18] since they can easily perfuse through the blood capillaries. The NPs can be delivered to the targeted site either via direct intratumoral infusion [19] or intravenously injected through cardiovascular network applying the effect of enhanced permeability and retention. Under the stimulation of external source (viz. infrared, magnetic, ultrasound and radiofrequency) [20, 21], the inorganic NPs (e.g. magnetic and gold NPs) are capable of absorbing and dissipating heat to the target tissue at a faster rate thereby enhancing the hyperthermia effect.

In order to improve the thermal ablation therapy, it is necessary to have a deep understanding of all the factors that can affect the tumor necrosis [22, 23]. Previously the tissue thermal field was mostly governed by the Pennes Bioheat diffusion equation [23] which consists of an external heat source term, metabolic heat generation and a blood perfusion term. It presumes a non-directional heat source term which accounts for the average heat sink effect due to the perfused blood through micro capillaries. Later on Wulff [24] claimed that the non-homogeneous effect of blood flow has to be governed as directional term in spite of scalar perfusion term. Hariharan et al [25] observed a noticeable change in tumor lesion size during HIFU treatment due to the heat sink effect of LBV. In an important work by Solovchuk et al. [26] have demonstrated the role of LBV coupling the thermal, acoustic and fluid field considering the acoustic streaming effect. Recently Vita et al. [27] had simulated the artificial vessels in an ex-vivo liver and observed a relevant tissue cooling adjacent to blood vessels up to 65%.

The classical Pennes bioheat equation is based on the assumption that the heat transfer in biological tissue occurs at an infinite speed which is practically not feasible as the thermal equilibration requires a finite time to occur. However this assumption of heat propagation at infinite speed is quite justified [28] for homogeneous medium with shorter time lag of $10^{-8}$ to $10^{-14}$ s. Micro anatomical study reveals that living tissue exhibits non-homogeneous nature [29] due to the presence of capillaries, cells, water content etc. in it. Thus the heat flux and temperature gradient occur at different instants of time. This would be very significant in case of HIFU treatment [28] especially when a very high temperature increment occurs at a very shorter period [30]. Therefore, a thermal relaxation time of heat flux ($\tau_q$) is incorporated in Pennes bioheat equation [28] to introduce the thermal wave model of bioheat transfer (TWMBT). The term $\tau_q$ represents the time taken to induce the temperature gradient in the tissue after imposing heat flux. Literature survey revealed different values of $\tau_q$, Mitra et al [31] and Roetzel et al [32] showed 16s and 1.77s respectively for processed meat whereas Li et al [33] have claimed the corresponding value of 5.66s for bovine liver tissue. With further modification in bioheat model, researchers explored a second time delay ($\tau_T$) which explains the incidence of micro structural interaction of complex bio tissue during heat transfer and is known as relaxation time due to temperature gradient. Taking into account both of the relaxation times ($\tau_q$ and $\tau_T$), dual phase lag (DPL) bioheat model is introduced by Tzou [34].

During thermal therapy, the non-uniform temperature field in tissue [35,36] can induce internal forces, resulting in thermomechanical deformation and thermal stress. This deformation can be governed by the thermoelastic wave equation. A minute alteration in mechanical stress induced under thermal load [35,36] on tissue can cause abnormality in hormone release and reduced immune response. Moreover the induced thermal stress can decrease the energy barrier [36] of acquiring protein denaturation boasting further the necrosis process of tissue. The tissue deformation during thermal treatment can further activate the pain sensing receptors to stimulate thermal pain [35, 36, 37]. Notably, Mcbride et al [37] developed a robust numerical model to explain the thermomechanical behavior during various therapeutic activities. Recently, Monteanthong et al [38] simulated the thermoelastic study on a breast cancer model during sonodynamic therapy and found that the therapeutic tissue displacement increases with increasing frequency of ultrasound.

The studies so far reported the thermal behavior of tissue during laser and ultrasound assisted cancer treatment in presence of vascular network and nanoparticles. However, to the best of author's knowledge no studies have been found related to thermomechanical analysis of vascularized tissue in presence of nanoparticles under the influence of focused ultrasound. Therefore in present numerical study an attempt has been made to compare the efficacy of LITT and HIFU therapy with pulsed and continuous mode of heating in terms of precise ablation of deeply seated vascularized tumor sparing the adjacent healthy tissue subjected to intravenous infusion of gold NPs. The tissue thermomechanical response has also been simulated to analyze the thermal pain induced in patients during cancer therapy. In present three-dimensional thermomechanical model, tissue inhomogeneity has been signified by incorporating LBV and two time lag terms ($\tau_q$ and $\tau_T$) for a breast carcinomas tumor. A preliminary experiment on agar based tissue phantom has been conducted to validate present numerical results of focused ultrasound heating.

## II. COMPUTATIONAL DETAILS

### A. Physical domain

A 3-dimentional multi-layered breast tissue model embedded in tumor and multilevel artery and vein (CVTT) was chosen as physical domain for present study and is demonstrated in Fig. 1. Here a conceptual demonstration of two thermal therapies (viz. HIFU and LITT) are shown together. To heat the tumor, the ultrasound wave was focused at the tumor location during HIFU therapy, whereas for LITT a thin catheter was modeled inside the tumor. In case of HIFU heating, the void space between the transducer surface and tissue was modeled as water to mimic the coupling medium [12]. To model the perfect absorption of sound at the surrounding walls, a perfectly matched layer (PML) of thickness 5 mm around the tissue was considered in the computational domain. The PML is not shown in Fig. 1 to avoid visual complexity. The focal depth (FD) from transducer bottom plane was taken as 50.67 mm [28]. The respective dimensions of computational domain are enlisted in Table: I (also shown in Fig. 1) according to literature [12, 36]. The branching of artery and vein in present study was made following the authors previous work [36]. The spacing between the artery and vein was kept as 5 mm [36]. The acoustic, optical, thermal and mechanical properties of the different layers of tissue medium were taken from literature [12, 39-44] and are enlisted in Table: II. A reference point P1 at a depth of 50.67 mm from tissue top surface and 5 reference planes viz. xy′, yz′, yz″, xz′, xz″ was considered for further studies and is shown in Fig. 1 with respective dimensions.

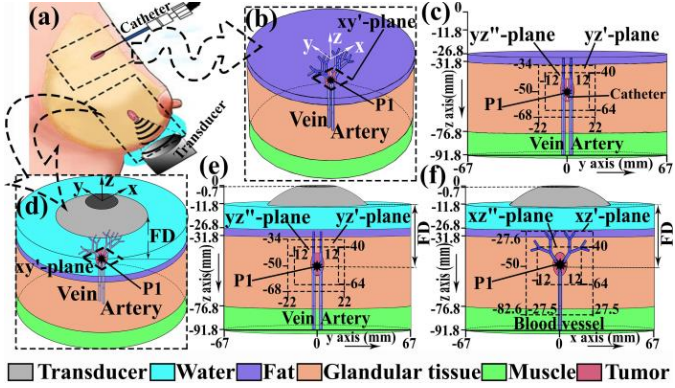

Fig. 1. (a) Conceptual demonstration of high intensity focused ultrasound (HIFU) and interstitial laser thermal therapy (LITT) of breast cancer; (b-f) conceptual 3-dimentional multilayered breast tissue model embedded in tumor and multilevel artery and vein during LITT (b, c) and HIFU therapy (d-f) with respective isometric (b, d) and side (c, e, f) views.

TABLE I
DIMENSIONS OF PRESENT COMPUTATIONAL DOMAIN [36, 12]

| Medium | Parameters |
|---|---|
| Fat | Radius=67mm, Height=5mm |
| Glandular tissue | Radius=67mm, Height=45mm |
| Muscle tissue | Radius=67mm, Height=15mm |
| Water | Radius=67mm, Height=15mm |
| Transducer | Radius=35mm, Height=11.8317mm |
| Catheter | Radius=0.1567mm, Height=13mm |
| Tumor | Major axis=14.7mm, Minor axis=4.5mm |
| Blood vessel (level 1) | Length=40mm, Radius=0.9mm |
| Blood vessel (level 2) | Length=11.5mm, Radius=0.7143mm |
| Blood vessel (level 3) | Length=8.131mm, Radius=0.567mm |
| Blood vessel (level 4) | Length=5.75mm, Radius=0.45mm |

$mm$ =millimeter

TABLE II
ACOUSTIC [12], OPTICAL [39-44], THERMAL [12] AND MECHANICAL [42, 43] PROPERTIES OF COMPUTATIONAL DOMAIN

| Medium | Values |
|---|---|
| **Acoustic Properties** | |
| Fat | $c_c = 1511, \alpha = 20.98$ |
| Glandular tissue | $c_c = 1588, \alpha = 106.55$ |
| Muscle tissue | $c_c = 1550, \alpha = 128.19$ |
| Tumor | $c_c = 1585, \alpha = 64.37$ |
| Blood | $c_c = 1540, \alpha = 15.32$ |
| Water | $c_c = 1483, \alpha = 0.488$ |
| **Optical Properties** | |
| Fat | $\mu_a = 227, \mu_s = 1030 (at.\lambda = 540)$ |
| Glandular tissue | $\mu_a = 358, \mu_s = 2440 (at.\lambda = 540)$ |
| Muscle tissue | $\mu_a = 116, \mu_s = 8820 (at.\lambda = 550)$ |
| Tumor | $\mu_a = 307, \mu_s = 1900 (at.\lambda = 540)$ |
| Blood | $\mu_a = 605, \mu_s = 3644 (at.\lambda = 560)$ |
| AuNp | $\mu_a = 67.8 \times 10^6, \mu_s = 45.9 \times 10^6 (at.\lambda = 550)$ |
| **Thermal Properties** | |
| Fat | $k = 0.21, C_p = 3000, \rho = 920, \omega_b = 0.18 \times 10^{-3}, Q_{met} = 400$ |
| Glandular tissue | $k = 0.48, C_p = 3000, \rho = 1080, \omega_b = 5.4 \times 10^{-4}, Q_{met} = 700$ |
| Muscle tissue | $k = 0.48, C_p = 3800, \rho = 1085, \omega_b = 5.4 \times 10^{-4}, Q_{met} = 700$ |
| Tumor | $k = 0.48, C_p = 3500, \rho = 1080, \omega_b = 0.009, Q_{met} = 34543.7$ |
| Blood | $k = 0.501, C_p = 4200, \rho = 1060$ |
| AuNp | $k = 317, C_p = 128, \rho = 19320$ |
| **Mechanical Properties** | |
| Fat | $E = 0.00416, \beta = 0.0001, \upsilon = 0.48$ |
| Glandular tissue | $E = 0.00385, \beta = 0.0001, \upsilon = 0.48$ |
| Muscle tissue | $E = 0.006, \beta = 0.0001, \upsilon = 0.48$ |
| Tumor | $E = 0.0102, \beta = 0.0001, \upsilon = 0.48$ |

$c_s$ =Sound speed in $m/s$, $\alpha$ =Attenuation coefficient at 1.5MHz in $dB/m$, $\mu_a, \mu_s$ =Absorption and scattering coefficient in $m^{-1}$ respectively at wavelength, $\lambda$ in $nm$, $k$ =Tissue thermal conductivity in $W/m \cdot K$, $C_p$ = Specific heat of tissue in $J/kg \cdot K$, $\rho$ =Density of tissue in $kg/m^3$, $\omega_b$ =Blood perfusion rate of tissue in $s^{-1}$, $Q_{met}$ =Metabolic heat generation rate in $W/m^3$, $E$ =Young's modulus in $MPa$, $\beta$ =Thermal expansion coefficient in $K^{-1}$, $\upsilon$ =Poisson's ratio

$m$ =meter, $s$ =second, $MHz$ =Mega Hertz, $dB/m$ =decibel per meter, $nm$ =nanometer, $W$ =Watt, $K$ =Kelvin, $J$ =Joule, $kg$ =kilogram, $MPa$ =mega Pascal

### B. Problem description

In the present study we have considered continuous and pulsed mode of wave in both LITT and HIFU thermotherapy. Basically three cases as mentioned in Table: III were compared in terms of thermal and mechanical response, considering the different operating loads, modes of energy imposition and infusion of gold nanospheres (AuNp). Here although a higher power and time was considered in case of C3 compared to C2, the applied activation energies in the respective cases are comparable since a very lower duty cycle (16.6%) was applied in the case of C3.

TABLE III
DIFFERENT CONSIDERED CASES OF ENERGY TRANSPOSITION MODES INFUSING AUNP INTRAVENOUSLY

| Conditions | Case 1 | Case 2 | Case 3 |
|---|---|---|---|
| AuNp addition with $\eta = 0.0002$ | No | Yes | Yes |
| Mode of energy exposure | Continuous | Continuous | Pulse |
| Time of energy exposure | $t = 10s$ | $t = 10s$ | $t = 150s$ |
| Power of energy exposure | $P = 1.6W$ | $P = 1.6W$ | $P = 2.6W$ |
| Duty cycle | -- | -- | $D = 16.6\%$ |
| Repetition time | -- | -- | $t_p = 0.3s$ |

$s$ = Second, $W$ = Watt

A comparison between different tissue models such as bare tissue (BT) and CVTT under HIFU and LITT heating in terms of thermal and mechanical response was also made. The light and acoustic field was governed by radiative transfer equation (RTE) and linearized pressure wave equation respectively. The fluid field was solved by momentum equation taking acoustic streaming effect. Besides, the thermal field was solved by both Pennes and dual phase lag (DPL) bioheat equations. To obtain the thermally induced stress, displacement, and strain field, equilibrium equation was solved. For simplification below assumptions were made.

1. Linear acoustic model was considered. As the applied focal pressure and intensity in this study were low, the non-linear propagation and cavitation can be neglected without having substantial errors [25, 45].
2. Constant optical, acoustic, thermal and mechanical property.
3. No phase change and chemical reaction in the tissue domain.
4. Laminar and incompressible blood flow.

### a. Thermal and fluid transport model

To compute the temperature history within the tissue both Pennes and DPL bioheat model was applied. The classical Fourier law based Pennes equation for tissue medium is written as below,

$$\rho C \frac{\partial T}{\partial t} = -\nabla \cdot q(r,z,t) + Q_{perf} + Q_{met} + Q_{ext}(r,z) \quad (1)$$

where,

$$q(r,z,t) = -k \nabla T(r,z,t) \quad (2)$$

where, $k$ is the thermal conductivity ($W/m \cdot K$), $C_p$ is the specific heat capacity ($J/kg \cdot K$), $\rho$ is the density ($kg/m^3$), $T$ is the temperature ($°C$), subscripts $b$ presents arterial blood, $\omega_b$ is blood perfusion rate ($s^{-1}$), $Q_{met}$ is volumetric metabolic heat generation ($W/m^3$), $Q_{ext}$ is volumetric external heat source ($W/m^3$), $Q_{perf}$ is the average volumetric heat exchange due to blood perfusion through capillaries and is given as,

$$Q_{perf} = \rho_b C_b \omega_b (T_b - T) \quad (3)$$

For considering the effect of thermal inertial and molecular level interaction, DPL bioheat model in 3-dimensional Cartesian coordinate system is given as,

$$q(x,y,z,t+\tau_q) = -k \nabla T(x,y,z,t+\tau_T) \quad (4)$$

By Taylors series expansion of Eq. (4) and putting the value of $\nabla \cdot q$ in Eq. (1), the final form of DPL model is expressed as,

$$\tau_q \rho C \frac{\partial^2 T}{\partial t^2} + (\tau_q \omega_b \rho_b C_b + \rho C) \frac{\partial T}{\partial t} = k \nabla^2 T + \tau_T k \nabla^2 \frac{\partial T}{\partial t} + \omega_b \rho_b C_b (T_b - T) \\ + Q_{ext} + Q_{met} + \tau_q \frac{\partial Q_{met}}{\partial t} + \tau_q \frac{\partial Q_{ext}}{\partial t} \quad (5)$$

where, $\tau_T$ and $\tau_q$ are the time delay terms ($s$) due to temperature gradient and heat flux respectively. In present study the value of $\tau_T$ and $\tau_q$ is considered as 0.043s and 16s respectively [46]. Constant values of $\tau_q$ and $\tau_T$ are assumed for the different layers of tissue.

The optical properties of uniformly doped tissue with AuNp were calculated using Eq. (6), during LITT [36, 47]. However, during HIFU therapy, the percentage of increase in tissue acoustic properties due to homogenous doping of AuNp during the numerical simulation were considered similar to measured value as shown in Table: IV. The thermal properties of AuNp doped tissue for both HIFU and LITT are calculated using Eq. (6) and (7) as given below.

$$M_{mix} = \eta M_{np} + (1-\eta) M_t \quad (6)$$

$$\frac{1}{k_{mix}} = \frac{\eta}{k_{np}} + \frac{1-\eta}{k_t} \quad (7)$$

where, $M$ is the tissue property, $\eta$ is the volumetric concentration of infused AuNp, here $e$ is the total quantity of AuNp per unit volume of tumor ($m^{-3}$) and $r_{np}$ is the radius ($nm$) of each AuNp,

$$\eta = e \left( \frac{4 \pi r_{np}^3}{3} \right) \quad (8)$$

The value of $\eta$ in present numerical study was taken as 0.0002 [11].

The normal body temperature of human body remains 37°C hence, initially; the temperature distribution within the tissue is considered to be uniform and is set as,

$$T_i(x,y,z,0) = 37°C \quad (9)$$

The interface of the different tissue layers were maintained as conservative flux condition, given as,

$$T_1 = T_2, k_1 \frac{\partial T_1}{\partial z} = k_2 \frac{\partial T_2}{\partial z}, z = l_1 \quad (10)$$

$$T_2 = T_3, k_2 \frac{\partial T_2}{\partial z} = k_3 \frac{\partial T_3}{\partial z}, z = l_1 + l_2 \quad (11)$$

Zero contact resistance between the different layers of tissue were considered,

$$n \cdot (k_1 \nabla T_1 - k_2 \nabla T_2) = 0, n \cdot (k_2 \nabla T_2 - k_3 \nabla T_3) = 0 \quad (12)$$

where, suffix 1, 2, 3 defines fat, glandular and muscle domain respectively and $l$ stands for thickness of tissue layers ($mm$). The other tissue surfaces were kept at constant temperature as,

$$T_{boundary}(x,y,z,t) = 37°C \quad (13)$$

The velocity and temperature distribution due to blood flow was governed by continuity, momentum and convective energy equation, given as,

$$\nabla \cdot v = 0 \quad (14)$$

$$\rho \left( \frac{\partial v}{\partial t} + v \cdot \nabla v \right) = -\nabla p + \nabla \cdot \Gamma + f_v \quad (15)$$

Paul *et al.*: Thermomechanical Assessment of Breast Tumor Subjected to Focused Ultrasound and Interstitial Laser Heating$$\rho C_b \frac{\partial T}{\partial t} + \rho C_b v \cdot \nabla T = \nabla \cdot (k \nabla T) + Q_{ext}(r,z) \quad (16)$$

where, $p, \rho, \Gamma, v, f_v$ are static pressure ($Pa$), density ($kg/m^3$), stress tensor ($N/m^2$), velocity ($m/s$) and volumetric force ($N/m^3$) respectively. The non-Newtonian blood flow was considered by assuming a power law and the expression for shear stress ($\xi$) is given as,

$$\xi_{xy} = b\left(\frac{dv}{dy}\right)^n = b\left|\frac{dv}{dy}\right|^{n-1}\frac{dv}{dy} \quad (17)$$

where, $b$ and $n$ are consistency and power law index respectively.

In the present study, the effect of acoustic streaming on velocity and temperature field was significant [25] since the blood vessels confined to the beam width (~3 mm). Therefore, the volumetric force, $f_v$ (Eq. (15)) acting on blood flow during HIFU was assumed to propagate along acoustic axis (z-axis). The non-zero component of $f_v$ along z axis is given as [25],

$$f_v = \frac{2\alpha_b}{c_b} I_s \quad (18)$$

where, $\alpha_b$ and $c_b$ are the acoustic attenuation coefficient ($dB/m$) and sound speed ($m/s$) of blood respectively, $I_s$ is the acoustic intensity ($W/m^2$). However, for LITT, $f_v$ is taken as 0.

The average velocity of blood at the artery inlet was taken as 0.03 m/s [36]. At the capillary bed, the same mass of blood from artery returns to the veins [36]. Hence in present study the blood was considered to be distributed equally into the 8 inflows of the small vein from 8 outlets of small artery. The artery inlet temperature of blood was taken as $37\,^\circ C$. The blood temperature at the inlet of veins was taken as surrounding tissue temperature since the capillary bed remains at thermally equilibrium condition. Thus,

$$T_{inlet.artery} = 37°C, T_{inlet.vein} = T_{tissue} \quad (19)$$

At blood vessel outlet pressure is given as Eq. (20), which is the average normal human blood pressure [36],

$$p = 80 mmHg \quad (20)$$

### b. Laser heating model

During LITT, the distribution of irradiated light energy within the tissue medium was governed by approximated radiative transfer equation (RTE) and is given as,

$$\nabla D \nabla \phi(r) + \mu_a \phi(r) = Q(r) \quad (21)$$

where, $\phi$ is the fluence rate of light energy ($W/m^2$), $Q$ is the volumetric source of injected power ($W/m^3$), $D$ is the coefficient of diffusion and is expressed as,

$$D = \frac{1}{3(\mu_a + \mu_s')} = \frac{\mu_a}{\mu_{eff}^2} \quad (22)$$

where, $\mu_a$ is the coefficient of absorption ($m^{-1}$), $\mu_s'$ is the reduced coefficient of scattering ($m^{-1}$), which is defined as,

$$\mu_s' = \mu_s(1-g) \quad (23)$$

where, $g$ is the factor of anisotropy that explains the directional dependency of scattering, $\mu_{eff}$ is the effective attenuation coefficient of light energy and is defined as

$$\mu_{eff} = \sqrt{3\mu_a(\mu_a + \mu_s')} \quad (24)$$

Thus the volumetric external heat source term ($Q_{ext}$) due to irradiation of laser light within the tissue medium as mentioned in above bioheat equations (Eq. (1) and (5)) is given as,

$$Q_{ext}(r,z) = \mu_a \phi(r) \quad (25)$$

In present study, the surrounding tissue surface emissivity is taken as 0.5.

### c. Ultrasound heating model

During HIFU therapy, the attenuated pressure energy was responsible for tissue heating. If $\alpha$ is the attenuation coefficient of acoustic energy then the source term ($Q_{ext}$) is written as,

$$Q_{ext} = 2\alpha I_s = 2\alpha \left| Re\left(\frac{1}{2}\rho_s v_s\right) \right| \quad (26)$$

where, $I_s$ is the ultrasound intensity ($W/m^2$), $p_s$ is the ultrasound pressure ($Pa$), $v_s$ is the ultrasound particle velocity ($m/s$). The acoustic pressure wave distribution within the tissue was governed by linear pressure wave equation, neglecting the viscous effect since the shear wave is very small in comparison to pressure wave. The general pressure wave equation neglecting density variation is given as [12],

$$\nabla \cdot \left[-\frac{1}{\rho_t}(\nabla p_s - \psi)\right] - \frac{1}{\rho_t c_s^2}\left(\frac{\partial^2 p_s}{\partial t^2}\right) = Q' \quad (27)$$

where, $c_s$ is the sound speed ($m/s$), $Q'$ and $\psi$ is the monopole and dipole source terms respectively. In present study $Q'$ and $\psi$ is assumed to be $0\ s^{-2}$ and $0\ N/m^3$. Considering these assumptions, the time dependent pressure wave equation can be expressed in frequency domain as,

$$\frac{1}{r}\frac{\partial}{\partial r}\left[-\frac{r}{\rho_t}\left(\frac{\partial p_s}{\partial r}\right)\right] + \frac{\partial}{\partial z}\left[-\frac{1}{\rho_t}\left(\frac{\partial p_s}{\partial z}\right)\right] - \frac{\omega^2}{\rho c_s^2}p_s = 0 \quad (28)$$

where, $\omega$ is the angular frequency ($rad/s$) and is defined as $2\pi f$, here $f$ is the frequency and is considered to be 1.5 MHz [12].

The power ($P$) of the acoustic wave in present study was related to the operating parameters of transducer as [28],

$$P = 2S\pi^2 \rho_W f^2 c_w \Delta x_{max}^2 = \frac{p_o^2 S}{2\rho_w c_w} \quad (29)$$

where, $S$ is transducer aperture area ($mm^2$), $\rho_w$ and $c_w$ are the density and sound speed of water (coupling medium between tissue and transducer), $\Delta x$ is the applied normal displacement ($nm$) of transducer aperture. $p_o$ is the source sound pressure ($Pa$) originated from transducer vibration.

### d. Formulation of pulse wave distribution

In present study, the distribution of electromagnetic and acoustic wave in pulsed mode within the tissue medium was followed by Gaussian temporal profile and is expressed as [12],

$$g(t) = \exp\{-4\ln 2 \times [((t - \frac{z}{c})/t_p) - 1.5]^2\} \quad (30)$$

where, $t$ is the time ($s$), $t_{pw}$ is the pulse width ($s$), $z$ denotes the depth from tissue top surface. Thus the resultant heat source term ($Q_{ext,pulse}$) in the above bioheat equations (Eq. (1) and (5)) for pulsed mode is given as,

$$Q_{ext,pulse} = Q_{ext,CW} \times g(t) \quad (31)$$

where, $Q_{ext,CW}$ is the external heat source term applied for continuous wave problem.

### e. Formulation of thermal damage

In order to quantify the extent of tumor necrosis, Arrhenius thermal damage ($\Omega$) was considered and is expressed as,

$$\Omega = \ln\left(\frac{C_0}{C_0 - C_d}\right) \quad (32)$$

where, $C_0$ and $C_d$ is the protein concentration of tumor before and after treatment respectively. The denaturation of protein happens at $\Omega \geq 0.53$ [48], whereas permanent denaturation of protein or cell death happens at $\Omega \geq 1$. Applying Arrhenius equation, the damage integral is calculated from the solution of bioheat equation. The damage [48] integral rate can be written as,

$$\frac{d\Omega}{dt} = A\exp(-G/UT(x,y,z,t)) \quad (33)$$

where, $G$ is the activation energy, considered as $6.28 \times 10^5 J/mol$, $T$ is the domain temperature ($K$), $U$ is the universal gas constant ($J/mol \cdot K$), $A$ is a pre-exponential factor ($3.1 \times 10^{98} s^{-1}$).

### f. Modeling of Thermoelastic deformation

Considering human tissue to be an elastic [36, 47], homogeneous and isotropic medium, the induced stress-strain relationship under thermal load for a 3-D Cartesian coordinate system can be expressed as,

$$\sigma_{xx} = \frac{E}{1+\upsilon}\varepsilon_{xx} + \frac{\upsilon E}{(1+\upsilon)(1-2\upsilon)}(\varepsilon_{xx}+\varepsilon_{yy}+\varepsilon_{zz}) - \frac{\beta E}{1-\upsilon}\theta \quad (34a)$$

$$\sigma_{yy} = \frac{E}{1+\upsilon}\varepsilon_{yy} + \frac{\upsilon E}{(1+\upsilon)(1-2\upsilon)}(\varepsilon_{xx}+\varepsilon_{yy}+\varepsilon_{zz}) - \frac{\beta E}{1-\upsilon}\theta \quad (34b)$$

$$\sigma_{zz} = \frac{E}{1+\upsilon}\varepsilon_{zz} + \frac{\upsilon E}{(1+\upsilon)(1-2\upsilon)}(\varepsilon_{xx}+\varepsilon_{yy}+\varepsilon_{zz}) - \frac{\alpha E}{1-\upsilon}\theta \quad (34c)$$

$$\sigma_{xy} = \frac{E}{1+\upsilon}\varepsilon_{xy}, \sigma_{xz} = \frac{E}{1+\upsilon}\varepsilon_{xz}, \sigma_{yz} = \frac{E}{1+\upsilon}\varepsilon_{yz} \quad (34d)$$

where, $\varepsilon$ is the normal strain, $\sigma$ is the normal stress, $\theta$ is the temperature difference between two consecutive steps of time, $E$ is the Young's modulus of elasticity, $\upsilon$ is the poison's ratio and $\beta$ is the coefficient of linear thermal expansion.

The strain $\varepsilon$ in terms of displacement vector ($u_x, u_y, u_z$) can be written as,

$$\varepsilon_{xx} = \frac{\partial u_x}{\partial x}, \varepsilon_{yy} = \frac{\partial u_y}{\partial y}, \varepsilon_{zz} = \frac{\partial u_z}{\partial z} \quad (35a)$$

$$\varepsilon_{xy} = \frac{1}{2}\left(\frac{\partial u_x}{\partial y}+\frac{\partial u_y}{\partial x}\right), \varepsilon_{xz} = \frac{1}{2}\left(\frac{\partial u_x}{\partial z}+\frac{\partial u_z}{\partial x}\right), \varepsilon_{yz} = \frac{1}{2}\left(\frac{\partial u_y}{\partial z}+\frac{\partial u_z}{\partial y}\right) \quad (35b)$$

The equations to govern the mechanics of non-rigid motion can be written as,

$$\rho\frac{\partial^2 u_x}{\partial t^2} = \frac{\partial \sigma_{xx}}{\partial x}+\frac{\partial \sigma_{xy}}{\partial y}+\frac{\partial \sigma_{xz}}{\partial z}+F_x \quad (36a)$$

$$\rho\frac{\partial^2 u_y}{\partial t^2} = \frac{\partial \sigma_{yx}}{\partial x}+\frac{\partial \sigma_{yy}}{\partial y}+\frac{\partial \sigma_{yz}}{\partial z}+F_y \quad (36b)$$

$$\rho\frac{\partial^2 u_z}{\partial t^2} = \frac{\partial \sigma_{zx}}{\partial x}+\frac{\partial \sigma_{zy}}{\partial y}+\frac{\partial \sigma_{zz}}{\partial z}+F_z \quad (36c)$$

where, $F$ is the body force under external condition. In present study $F$ is taken as 0.

To solve the induced thermal stress, the predicted temperature (Eqs. (1) and (5)) were taken as input into the thermo-elastic model (Eq.(34)). After that by solving the equations of non-rigid motion mechanics (Eq.(36)), the thermomechanical displacement was computed. Finally, by summing up equations (34)-(36), the thermally induced displacement can be written as,

$$\rho\frac{\partial^2 u_x}{\partial t^2} = \frac{E}{2(1+\upsilon)}\left(\frac{\partial^2 u_x}{\partial x^2}+\frac{\partial^2 u_x}{\partial y^2}+\frac{\partial^2 u_x}{\partial z^2}\right) + \frac{E}{2(1+\upsilon)(1-2\upsilon)}\frac{\partial}{\partial x}\left(\frac{\partial u_x}{\partial x}+\frac{\partial u_y}{\partial y}+\frac{\partial u_z}{\partial z}\right) - \frac{E\beta}{1-2\upsilon}\frac{\partial\theta}{\partial x}+F_x \quad (37a)$$

$$\rho \frac{\partial^2 u_y}{\partial t^2} = \frac{E}{2(1+\upsilon)} \left( \frac{\partial^2 u_y}{\partial x^2} + \frac{\partial^2 u_y}{\partial y^2} + \frac{\partial^2 u_y}{\partial z^2} \right)$$
$$+ \frac{E}{2(1+\upsilon)(1-2\upsilon)} \frac{\partial}{\partial y} \left( \frac{\partial u_x}{\partial x} + \frac{\partial u_y}{\partial y} + \frac{\partial u_z}{\partial z} \right) - \frac{E\beta}{1-2\upsilon} \frac{\partial \theta}{\partial y} + F_y \quad (37b)$$

$$\rho \frac{\partial^2 u_z}{\partial t^2} = \frac{E}{2(1+\upsilon)} \left( \frac{\partial^2 u_z}{\partial x^2} + \frac{\partial^2 u_z}{\partial y^2} + \frac{\partial^2 u_z}{\partial z^2} \right)$$
$$+ \frac{E}{2(1+\upsilon)(1-2\upsilon)} \frac{\partial}{\partial z} \left( \frac{\partial u_x}{\partial x} + \frac{\partial u_y}{\partial y} + \frac{\partial u_z}{\partial z} \right) - \frac{E\alpha}{1-2\upsilon} \frac{\partial \theta}{\partial z} + F_z \quad (37c)$$

The lower surface of the tissue domain is given as fixed boundary condition whereas the rest of the boundaries are maintained free. The initial (before heating) stress, strain values are taken as 0.

$$\sigma_{xi}, \sigma_{yi}, \sigma_{zi}, \sigma_{xyi}, \sigma_{xzi}, \sigma_{yzi}, \varepsilon_{xi}, \varepsilon_{yi}, \varepsilon_{zi}, \varepsilon_{xyi}, \varepsilon_{xzi}, \varepsilon_{yzi} = 0 \quad (38)$$

### C. Numerical simulation and validation

In the present numerical study, the linearized pressure wave, momentum, RTE, bioheat, Arrhenius and equilibrium equations were solved by finite element based COMSOL$^{TM}$ Multiphysics (COMSOL Inc, Bengaluru, India) commercial software. During the simulation, both the relative and absolute tolerance was maintained as 0.001. The model was discretised by quantic Lagrange method for acoustic physics whereas quadratic Lagrange method was applied for the rest of the physics. The time steps taken by the solver was made "strict" using Implicit Backward Differential formula with a maximum time step of 0.05 s. The mesh convergence test of the present model was performed with various element numbers as shown in Fig. 2 (a). A mesh element of 3,16,000 was chosen for the entire study. The time step independency test (Fig. 2 (a)) was also performed and subsequently a time step of 0.05 s was chosen. The simulation was performed in Dell Precision Tower 7810 (Intel Xeon E5 v4 Processor, 20 Core, 96GB DDR4 RAM).

The present numerical results were verified with the known solutions available in the literature [5, 12, 38, 49]. The acoustic pressure field was compared with the results of Bhowmik et al. [12] (Fig. 2 (b)). The temperature field during the continuous and pulsed mode heating of tissue were compared with the results of Bhowmik et al. [12] (Fig. 2 (c)) and Ganguly et al. [5] (Fig. 2 (d)) respectively. The temperature field during interstitial laser heating was compared with Marqa et al. [49] (Fig. 2 (e)). The mechanical field in terms of thermally induced tissue displacement was compared with Montienthong et al. [38] (Fig. 2 (f)). A good agreement was obtained between the present numerical simulation and previous studies. For validation of the simulation results, all the operating parameters, properties and boundary conditions were kept identical to previous study. Furthermore, the simulation results of focused ultrasound heating were also validated experimentally with equivalent tissue phantoms and are discussed in section IV.A.

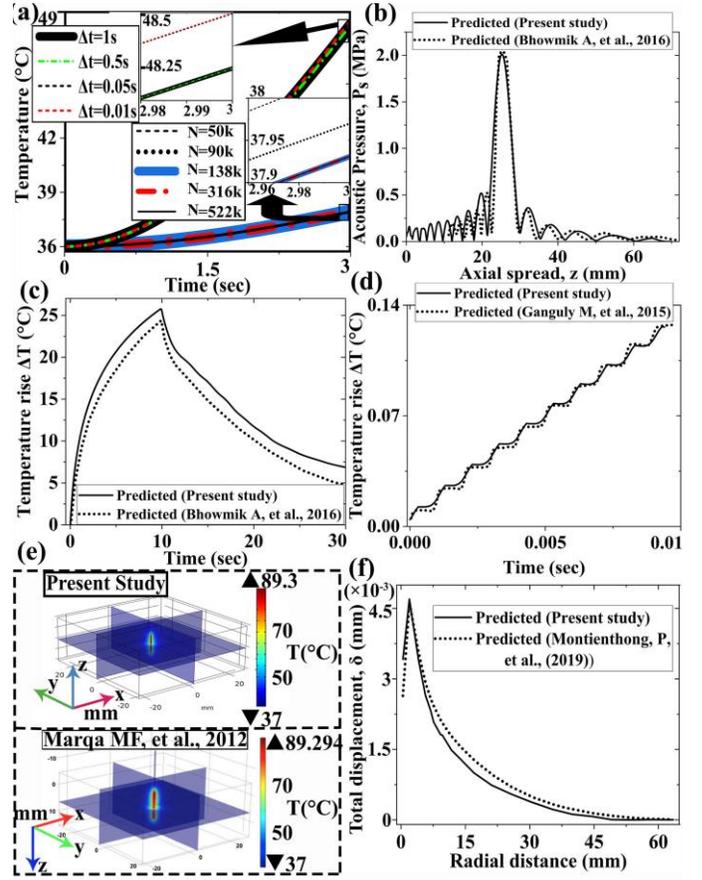

Fig.2 (a) present temporal temperature plot of tissue at various time step and grid size; (b-f) comparison of present and benchmark results viz. (b) acoustic pressure distribution over z-axis at t=10s, (c, d) tissue temperature rise over time during continuous (c) and short pulsed (d) mode sonication of ultrasound, (e) spatial temperature distribution at t=125s during interstitial laser heating (f) tissue displacement history over radial distance for 10s sonication of ultrasound

### III. EXPERIMENTAL METHODS

#### A. Preparation of nanoparticle mixed tissue phantom

The base material chosen for preparing the tissue phantom was agar powder which was brought from Sisco Research laboratories (Mumbai, India). The procedure of making the agar gel was same as author's previous work [36]. To prepare the nanoparticle mixed tissue phantom, the gold nanospheres (d<100nm) in powder form was brought from Sisco Research laboratories (Mumbai, India). The gold nanosphere (AuNp) suspension in distilled water was prepared by sonicating the mixture for 5-10 minutes. The nanoparticle solution then gently mixed with the agar gel solution and stirred continuously using a magnetic stirrer (TARSONS [Kolkata, India]) to produce a homogeneous mixture of agar gel and AuNp. The nanoparticle mixed agar gel was poured into the test acrylic basin and kept undisturbed for 12 hours to solidify.

#### B. Measurement of thermal and acoustic properties

The thermal properties of tissue phantom were measured using the thermal property analyser (TEMPO, Meter Group, Inc., USA). The acoustic properties of the tissue phantom were measured following the substitution method [50]. This method is basically based upon the comparison of ultrasound signal strength transferred through the tissue phantom with respect to controlled medium (water). The setup for measuring the acoustic properties is demonstrated in Fig. 3 (A). 1-MHz transducer (50 inch focus, 20 mm aperture, Roop Telsonic Ultrasonix Limited, Mumbai India), driven by a frequency generator (FY6900-60M, FeelTech, Henan Province, China) at 24 V was faced a confocally aligned receiver transducer of capacity 2.5 MHz (50 inch focus, 20 mm aperture, Roop Telsonic Ultrasonix Limited). The phantom was centered coaxially between the two transducers, submerged in a distilled water tank. To measure the attenuation, the emitter transducer was signalled with a 30-cycle sineburst at 1 MHz frequency from frequency generator with and without tissue phantom. Then the received voltage signal was closely monitored in a digital storage oscilloscope (SMO120E, Scientific Mes Technik Private Limited, MP, India). To measure the sound speed, the emitter transducer was signalled by a pulse generator (RTUL MHF-400, Roop Telsonic Ultrasonix Limited, Mumbai India) with and without the phantom. Thereafter the output signal from the receiver transducer was recorded with the digital storage oscilloscope.

Based on the assumption of exponential decay, the attenuation coefficient, $\alpha$ of the tissue phantom was calculated using the Eq. (36) [50], with known sample thickness $d$ as,

$$\alpha = 20 \frac{1}{d} \log(V_w / V_p) \quad (39)$$

where, $V_w$ and $V_p$ are the voltage amplitude of the transmitted signal through water in absence and presence of tissue phantom respectively.

Whereas, the speed of sound, $c_s$ through the tissue phantom was calculated [50] as,

$$c_s = d / \{(d / c_w) + \Delta t\} \quad (40)$$

where, $c_w$ is the sound speed through water, $\Delta t$ is the time shift in received pulse when the phantom was submerged in water. The time shift was measured and analyzed in MATLAB to calculate the speed of sound.

The measured thermal and acoustic properties of tissue samples are enlisted in Table: IV.

TABLE IV
MEASURED THERMAL AND ACOUSTIC PROPERTIES OF AUNP DOPED TISSUE PHANTOM AT DIFFERENT CONCENTRATIONS

| Tissue Phantom | Without doped ($\eta = 0$) | AuNp doped ($\eta=0.000045$) | AuNp doped ($\eta=0.0002$) |
|---|---|---|---|
| Sound speed, $c_s$ in $m/s$ | 1500±0.02 | 1507.35±0.01 | 1532.7±0.03 |
| Acoustic attenuation coefficient, $\alpha$ in $dB/m$ | 6.08±0.017 | 6.688±0.04 | 36.8±0.03 |
| Specific heat, $C_p$ in $J/kg \cdot K$ | 3913±45.4 | 3912.8±42.1 | 3912.24±43.3 |
| Thermal Conductivity, $k$ in $W/m \cdot K$ | 0.6±0.003 | 0.6±0.006 | 0.6±0.005 |
| Density, $\rho$ in $kg/m^3$ | 1050±0.018 | 1050.8±0.02 | 1053.65±0.03 |

$s$ =second, $m$ =meter, $dB$ =decibel, $J$ =Joule, $kg$ =kilogram, $K$ =Kelvin, $W$ =Watt

### C. Procedure for ultrasound heating experiments

The full setup of experiment is shown in Fig. 3 (B). The setup consists of an ultrasound therapy system (LCS-128, Life Care Systems, Uttar Pradesh, India), an acrylic test section containing agar phantom, K-type thermocouples for temperature measurement, a data logger (34980A, Keysight Technologies Private Limited, Bangalore, India), and a computer for monitoring. The transducer of the therapy system was signalled with 1 MHz frequency at a power of 15 W. The system was turned on for 214 s for heating and turned off subsequently for cooling for another 325 s. The transducer was held in accurate position with a custom made stand. The height and diameter of the acrylic test section were 15 cm and 10 cm respectively with arrangements for inserting the thermocouples radially. Ultrasound gel was provided between the tissue phantom and ultrasound probe as a coupling agent. The K-type thermocouples were located in the focal zone to scan the temperature in real time.

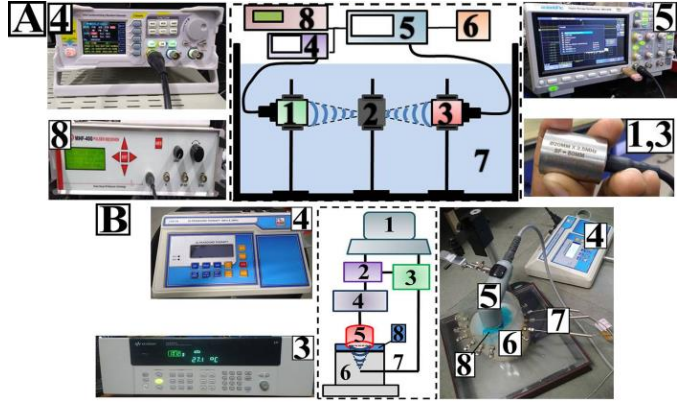

Fig 3 (A) Acoustic property measurement setup consisting of (1) ultrasound transducer (emitter), (2) tissue phantom, (3) ultrasound transducer (receiver), (4) Sineburst frequency generator, (5) digital storage oscilloscope, (6) power source, (7) distilled water tank, (8) pulse frequency generator; (B) Ultrasound assisted tissue heating setup consisting of (1) computer, (2) power source, (3) data logger, (4, 5) ultrasound therapy system, (6) tissue phantom, (7) thermocouples, (8) ultrasound gel.

## IV. RESULTS

### A. Comparison of measured and predicted results of ultrasound heating

A preliminary experiment was carried out on agar based tissue phantom to validate the present numerical results of ultrasound assisted heating. The measured and predicted temporal temperature plot at a distance of 17.5 mm from the top surface of phantom is shown in Fig. 4. The location of temperature measurement was considered within the range of ultrasound focal zone. The experiment was conducted on two tissue phantoms, (a) without AuNp doped, and (b) with AuNp doped ($\eta=0.000045$). In this section the properties taken in simulation were measured and are listed in Table. IV. The trend of measured temperature profile well agreed with the numerical results both qualitatively and quantitatively with an error of 6.2% and 11.4% for without and with nanoparticle mixed tissue phantom respectively.

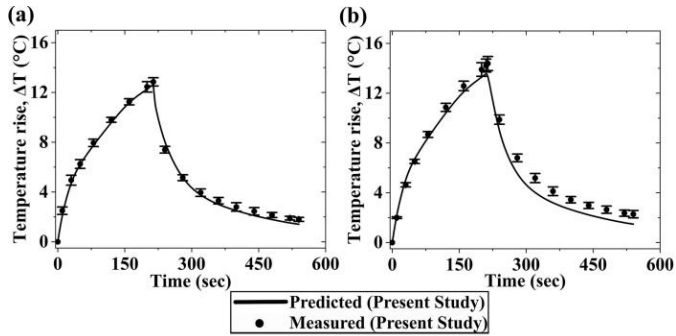

Fig.4 Comparison of measured and predicted temporal thermal history of tissue phantom without nanoparticles (a) and with nanoparticles (b) at a point (17.5 mm below the top surface) during the ultrasound heating (at 1 MHz, 15 W).

### B. Tissue thermal response under inhomogeneous condition

Figure 5 depicts the tissue (without and with LBVs) thermal and damage responses predicted by Pennes and DPL thermal model. Specifically, the temperature and damage distribution over time at point P1 for the cases of C1 (without NPs, continuous mode), C2 (with NPs, continuous mode) and C3 (with NPs, pulsed mode) (as given in Table III) considering both Pennes and DPL model within the BT and CVTT model is shown. During the HIFU treatment, the maximum tissue temperature as predicted by Pennes model has been found to be 85°C and 7°C higher in comparison to DPL model for the cases of C2 and C3 respectively. However during LITT, the corresponding higher temperature prediction by the Pennes model has been found to be 39°C and 25°C, respectively within the tissue in absence of LBVs (Fig. 5 (a) and (c)). Thus, the DPL model predicted a different tissue thermal history than the classical Fourier based Pennes bioheat model for the same operating parameters. For HIFU heating, the maximum tissue temperature rise at P1 reached to 96°C for both cases of C2 and C3 with a sonication time of 10 s and 150 s and power of 1.6 W and 2.6 W respectively. However for LITT, the maximum rise in tissue temperature reached to 85°C for both the cases of C2 and C3 in Fig. 5 (a) and (c). Further, the presence of LBVs (CVTT) reduced the maximum rise in tissue temperature by 24°C compare to BT for both HIFU and LITT considering the DPL model (Fig. 5 (a) and (c)).

The trend of Arrhenius damage integral with respect to time was calculated based on the predicted temperature distribution for all the considered cases as shown in Fig. 5 (b) and (d). Noticeably, the time taken for achieving the threshold necrotic damage ($\Omega \geq 1$) has been found to be different for all of the considered cases (Fig. 5 (b) and (d)). With the same operating loads (P=1.6 W, t=10 s), an over prediction in temperature by 51°C has been observed for C2 relative to C1 during HIFU whereas the same for LITT has been found to be 45°C as shown in Fig. 5 (a) and (c). Moreover, at point P1 and t=10 s (DPL model), the $\Omega$ value is predicted to be $4\times10^{-4}$ and $4\times10^{-6}$ for C1 in HIFU and LITT, respectively. The corresponding values of $\Omega$ for C2, is $7.2\times10^{8}$ and $1.3\times10^{3}$ respectively, as shown in Fig. 5 (b) and (d). Interestingly, for the case C1, the value of $\Omega$ failed to reach the necrotic limit ($\geq 1$) even after 10 s of exposure, whereas for C2, the threshold limit of $\Omega$ has been achieved within 4.9s and 2.6s for HIFU and LITT respectively.

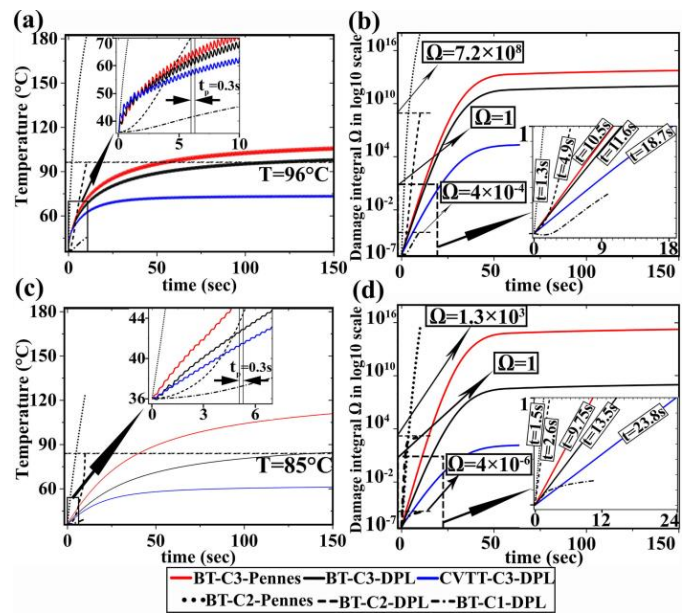

Fig.5 Prediction of tissue temperature (a, c) and damage (b, d) history over time at P1 for the ultrasound (HIFU) (a, b) and laser (LITT) (c, d) based heating of tissue with different modes (viz. C1, C2 and C3) for different tissue models (viz. BT, CVTT) and bioheat models (viz. Pennes, DPL).

### C. Effect of pulsed energy on necrotic temperature and damage distribution

The spatial temperature distribution on yz′-plane (Fig. 1) for the cases of C1, C2, and C3 is shown in Fig. 6. In the following sections DPL bioheat model has been applied. During ultrasound heating of tissue in absence of LBVs, a wider necrotic temperature zone (T≥52°C) [36, 47] has been depicted in case of nanoparticle mixed tissue with pulsed energy activation (Column 3, Row 1) compared to the continuous mode (Column 2, Row 1). The necrotic zone is confined to the entire tumor regime preserving the surrounding normal tissue in case of C3 compared to C2. A narrower necrotic zone with further reduction in peak temperature (T=72°C) has been predicted in case of CVTT compared to BT (Column 3, Row 2 and 1).However, during laser heating of bare tissue, the necrotic zone spreads to the healthy regime in case of C3 compared to C2 (Fig. 6 [Column 3 and 2, Row 3]). Whereas, the expansion of necrosis zone is limited within the tumor boundary in case of C3 for CVTT (Column 3, Row 4) with a decrease in peak temperature (T=61°C) compared to BT (Column 3, Row 3). Notably, with constant operating loads, the necrotic temperature profile is found in case of C2 (with NPs) which is absent in case of C1 (without NPs) as shown in Fig. 6.

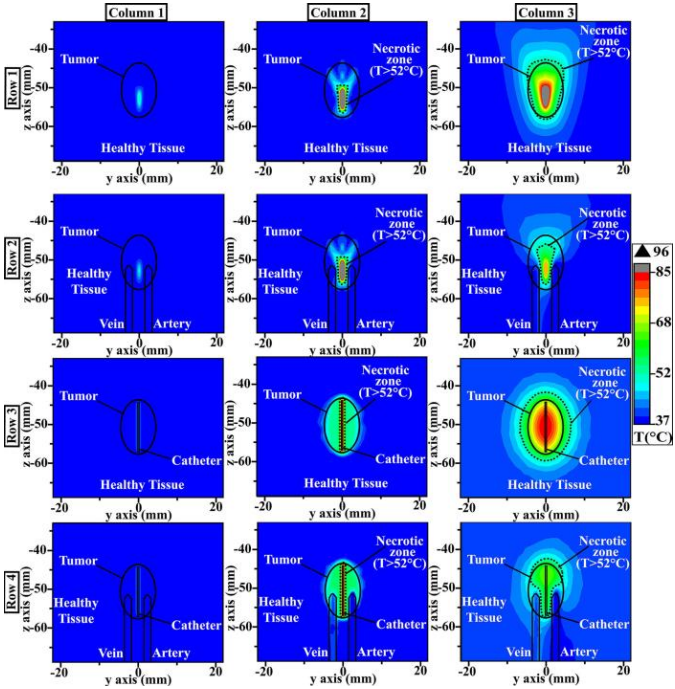

Fig.6 Spatial temperature contour on yz'-plane for the cases of C1 at t=10 s [Column 1], C2 at t=10 s [Column 2] and C3 at t=150 s [Column 3] considering BT [Row 1 and 3] and CVTT [Row 2 and 4] during HIFU [Row 1 and 2] and LITT [Row 3 and 4].

Fig. 7 shows the spatial necrotic damage ($\Omega \geq 1$) [48] distribution on different planes of tissue for various modes of energy activation and treatment methods. A wider necrotic damage distribution has been predicted in case of C3 compared to C2 for all the considered cases. Noticeably, no necrotic damage history has been found in case of tissue in absence of nanoparticles (C1) while keeping the same operating parameter as C2. However, during pulsed mode of energy activation, the entire tumor volume has been predicted to be irreversibly damaged sparing the healthy tissue more precisely during HIFU therapy (Row 1) compared to LITT (Row 3) for all considered planes of BT model. The corresponding necrotic damage distribution was predicted to be less in case of tumor tissue in presence of LBVs compared to BT model (Fig. 7 [Row 4 and 2]).

### D. Effect of multilevel artery and vein on ultrasound and laser heating

The spatial temperature distribution on xz′-plane for different treatment conditions are displayed in Fig. 8. During the LITT with pulsed mode of energy supply, heat diffusion took place at the top half of the tissue in the vicinity of artery whereas, the bottom half of the tissue gets heated due to the presence of vein (Row 1 and 3). In case of continuous mode, the heat is confined mostly to the tumor region (Row 2 and 4). Moreover, the effects of artery and vein have been found to be absent in terms of qualitative thermal history during ultrasound heating in both the cases of C2 and C3 (Row 3 and 4).

### E. Mechanical response of multi-layered tissue

Fig. 9 illustrates the stress-strain curve of tissue at location P1 under various energy deposition modes and treatment methods. A non-linear stress-strain profile has been predicted for all considered cases. Besides a higher stress-strain value is predicted in case of nanoparticle mixed tumor compared to the tumor in absence of nanoparticles. Moreover, for an applied strain a lower stress is induced in case of pulsed energy activation compared to continuous mode for all the considered tissue models and treatment methods as shown in Fig. 9. A hysteretic stress-strain profile with dissimilar loading and unloading path has been observed during pulsed mode of energy activation for both the external heating. Interestingly, the area under the corresponding hysteresis loop is reduced for the CVTT (Fig. 9 (b), (d)) compared to BT (Fig. 9 (a), (c)) model for both the treatment methods. A dissimilar stress-strain profile for the cases C1, C2 and C3 modes with a reduced hysteresis loop size particularly for C3 mode has been observed in case of LITT (Fig. 9 (c), (d)) compared to HIFU (Fig. 9 (a), (b)) therapy.

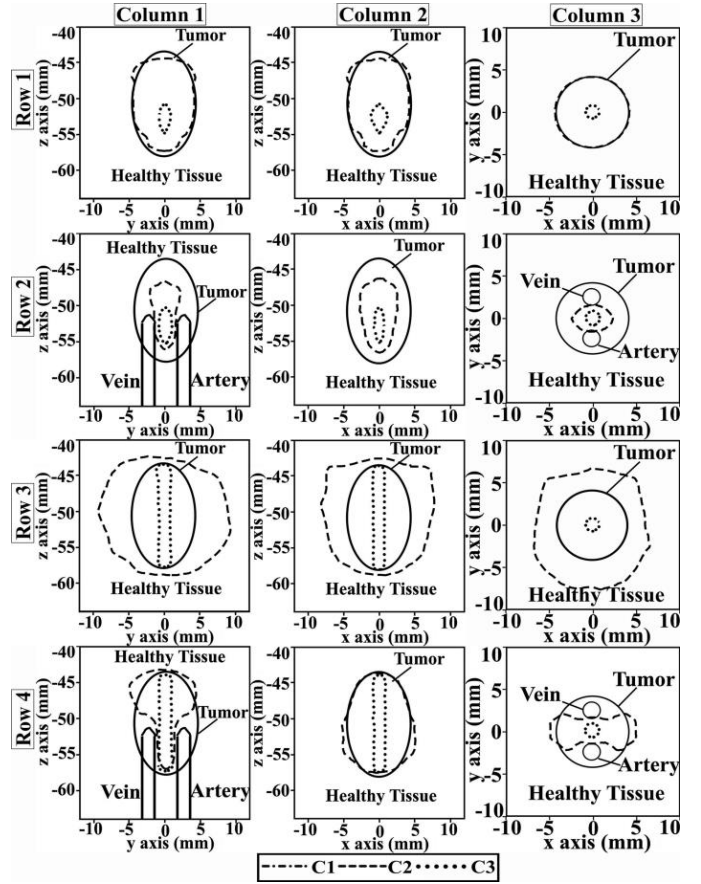

Fig.7 Spatial necrotic damage plot for the cases of C2 at t=10s and C3 at t=150s on yz''-plane [Column 1], xz''-plane [Column 2], xy'-plane

[Column 3], considering BT [Row 1 and 3] and CVTT [Row 2 and 4] model under HIFU [Row 1 and 2] and LITT [Row 3 and 4].

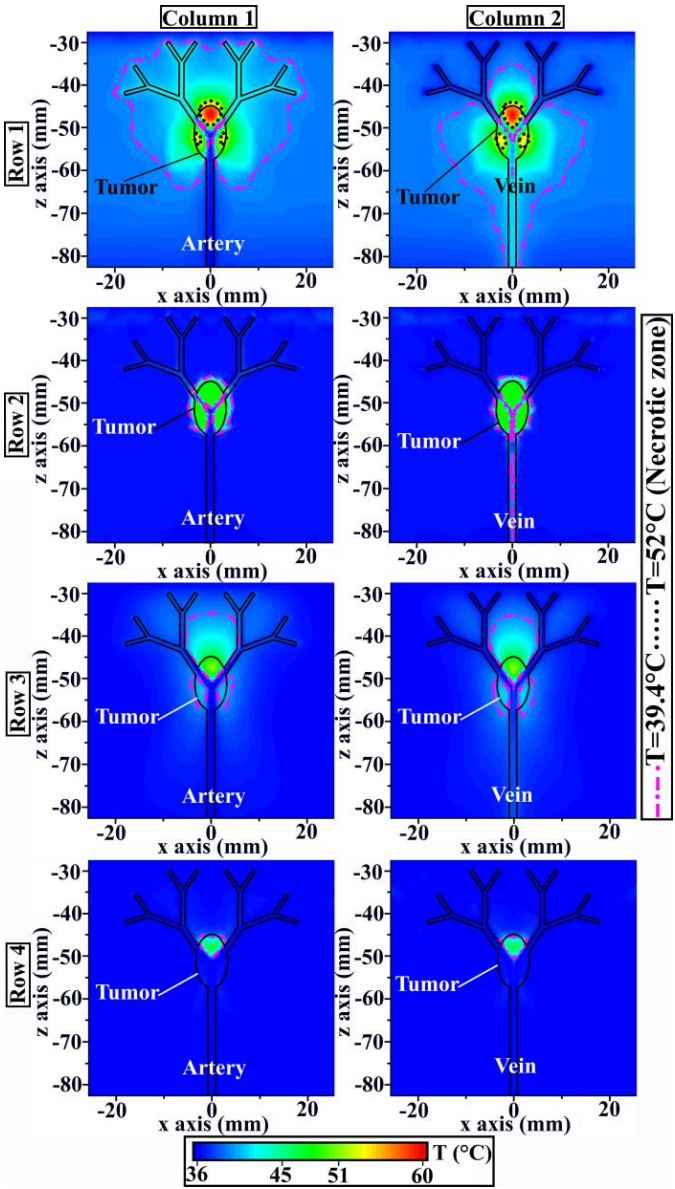

Fig.8 Spatial temperature distribution on xz'-plane across the artery [Column 1] and vein [Column 2] for the cases of C3 at t=150s [Row 1 and 3] and C2 at t=10s [Row 2 and 4] during LITT [Row 1 and 2] and HIFU therapy [Row 3 and 4].

Fig. 10 shows the induced stress history of tissue under thermal load on yz''-plane (Fig. 1) for different modes of energy deposition considering various tissue models and treatment methods. The stress contour of magnitude 18.3 Pa expands over a wider area for C2 compared to C3 for all presented tissue models and treatment methods (Column 2, 3). The corresponding stress contour has been found to be absent for C1 (Column 1). The thermal stress profile of 77.6 Pa is predicted only in case of C2 for both BT and CVTT model (Column 2, Row 1, 2). The span of thermal stress (18.3 Pa) within the tumor has been found to be reduced in case of CVTT compared to BT for pulsed HIFU heating (Column 3, Row 1, 2). However, the corresponding stress history is seen only in case of CVTT unlike BT during LITT (Column 3, Row 3, 4). Besides, almost a similar stress profile (18.3 Pa) is witnessed for both the BT and CVTT under C1 and C2 mode of heating (Column 1, 2). On the other hand, a reduced stress history (18.3 Pa) has been observed for LITT (Row 3, 4) in comparison to HIFU therapy (Fig. 10 [Row 1, 2]) with fixed operating loads.

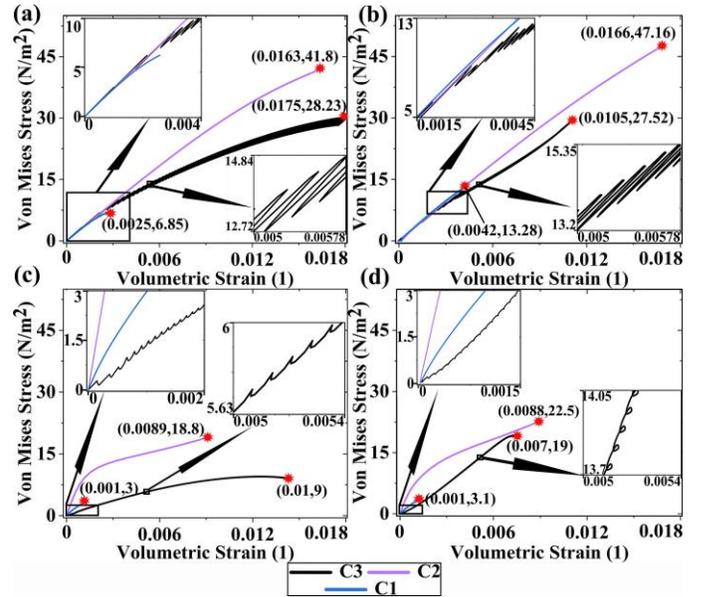

Fig.9 Viscoelastic stress-strain response of tissue at P1 under thermal load for the cases of C1, C2 and C3 considering different models viz. (a) HIFU in BT, (b) HIFU in CVTT, (c) LITT in BT, and (d) LITT in CVTT.

The tissue deformation and displacement history under various heating modes and treatment methods is presented in Fig. 11 for different tissue models. The maximum tissue displacement under the thermal load for each considered case is also shown. The tissue thermal deformation is found to be highest in case of C3 successively after C2 and C1. Besides, a reduced thermally induced tissue deformation has been revealed in case of CVTT comparing BT during both HIFU therapy and LITT as shown in Fig. 11. However, the tissue has deformed more under thermal load for LITT compared to HIFU. Notably, the tissue displacement is revealed more above the tumor regime compared to other regions of computational domain. Indeed, the corresponding deformation was found to be dispersed in case of LITT compared to HIFU therapy.

## V. Discussions

In this study, the thermal and mechanical response of tissue under ultrasound and laser exposure was analyzed numerically to understand the effects of tunable parameters on precise tumor ablation avoiding damage to the surrounding healthy tissue with minimum nociceptive pain. We have compared the results of pulsed and continuous mode energy deposition during HIFU and LITT considering both Pennes and DPL bioheat model for different operating loads as well as with and without AuNp. Further, tissue models without blood vessels (BT) and with countercurrent multilevel blood vessels (CVTT) were considered. Additionally, the non-homogeneous nature of tissue was considered by the tissue thermal relaxation times ($\tau_q$, $\tau_T$) and multilevel arteries and veins embedded in triple-layered tissue structure. An *in-vitro* study on agar based tissue phantom was also performed to validate the present numerical results for ultrasound therapy.

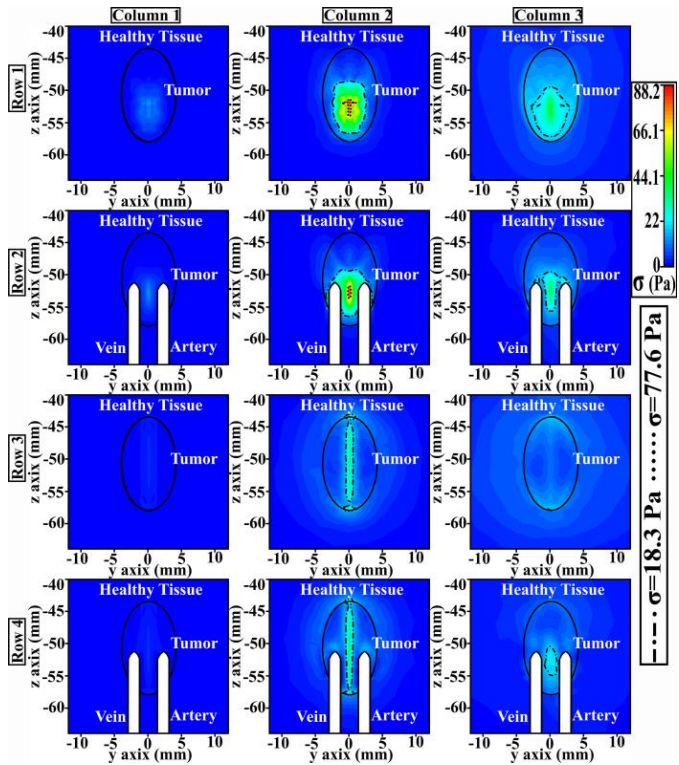

Fig.10 Spatial distribution of induced thermal stress on yz"-plane of BT [Row 1 and 3] and CVTT [Row 2 and 4] model considering the cases of C1 at t=10s [Column 1], C2 [Column 2] and C3 [Column 3] during HIFU therapy [Row 1 and 2] and LITT.

In the DPL model, an initial delay in steep rise of temperature was observed due to the non-Fourier effects ($\tau_q$, $\tau_T$) of the bio-tissues, which defines the finite speed of thermal wavefront as well as the delay in microstructural interaction of the tissue. Interestingly, during HIFU heating, the difference in temperature prediction between Pennes and DPL model has been revealed more (78 °C) for continuous mode heating compared to pulse mode (Fig. 5(a)). This is expected because during focused ultrasound heating with continuous mode, the temperature rises rapidly at the focal point P1 in comparison to the pulsed mode while the Pennes model predicts the tissue thermal history by assuming an infinite speed of the thermal wavefront. However during the unfocused laser heating compared to focused ultrasound heating, the difference in temperature prediction between Pennes and DPL model was found to be slightly more (14 °C) for continuous heating in respect of pulse heating (Fig. 5 (c)). Therefore, during focused ultrasound heating, especially for shorter time period, it is essential to employ the DPL bioheat model for the prediction of accurate tissue temperature history.

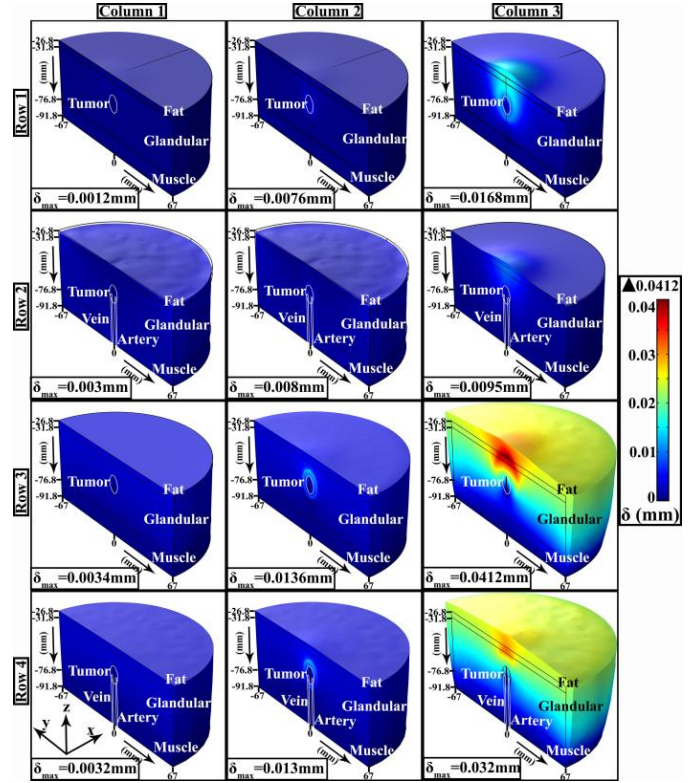

Fig.11 3-dimensional tissue deformation with displacement history (symmetric half portion)) for C1 [Column 1], C2 [Column 2], and C3 [Column 3] in the cases of BT [Row 1, 3] and CVTT [Row 2, 4] model during HIFU therapy [Row 1, 2] and LITT [Row 3, 4].

The tissue temperature at the targeted zone should be in the range of 60 °C and 95 °C for protein coagulation or necrosis of tumor cell [36, 51]. The effective temperature gradient in hysteretic manner for pulsed heating is found to be less in comparison to continuous heating. The cooling effects of LBVs along with the microcapillaries make the present bioheat model more realistic in terms of modeling inhomogeneous tissue media (Fig. 5 (a) and (c)). Interestingly, the time taken by the tissue to attain the necrotic damage ($\Omega \geq 1$) varies under different heating conditions as shown in Fig. 5 (b) and (d). The focused ultrasound heating causes a higher temperature rise in tissue compared to the unfocused laser heating as shown in Fig. 5 and 6. During the pulse heating with lower duty cycle (16.6%) the tissue absorbs the equivalent activation energy for longer period compared to continuous heating (Table: III) which affects the respective necrotic temperature span (Fig. 6 [Column 3 and 2]). Thus, the pulsed mode energy deposition with lower duty cycle worked more effectively during the focused energy source (HIFU) as well as the unfocused energy source (LITT).

The extent of tumor damage occurred under various modes of energy activation during HIFU therapy and LITT have been justified by the corresponding necrotic damage ($\Omega \geq 1$) plot across different planes (Fig. 7). The enhanced thermal response of tissue with uniform mixing of nanoparticles improves the necrotic damage history (Fig. 6 and 7). Moreover, the pulsed mode of energy activation along with lower duty cycle for longer period improves the necrotic damage distribution compared to the continuous mode. It is also observed that the heat dissipating effect of counter current blood vessels reduces the effective span of necrotic damage history. During interstitial laser heating, the blood flow through artery and vein with initial body temperature distribute the accumulated heat of tumor to the nearby healthy tissue (Fig. 8 [Row 1]). Indeed the role of artery-vein on tissue heating has been found to be more prominent when the tissue is exposed to the heating for longer period (Fig. 6 and 8).

The stiffness of viscoelastic biological medium affects the thermally induced tumor necrosis [52]. The time dependent elastic moduli of present stress-strain curve reflect the viscoelastic behavior of tissue during thermal therapy. The enhanced thermal response of tissue with homogeneous doping of nanoparticles results in an increased strain rate. Besides with limited peak tissue temperature, the stress-strain curve suggests a lower strain rate and stiffness of tissue medium under pulse heating compared to continuous heating under present condition (Fig. 5, 9). The area under the hysteresis loop of stress-strain curve defines the absorbed energy of a viscoelastic medium after each loading and unloading phase of the pulsed heating. Interestingly, some amount of this absorbed energy is dissipated by the counter current blood flow thus, reducing the effective area under each hysteresis loop as shown in Fig. 9. As a result, the effective rise in strain with respect to stress after each cycle is reduced for CVTT model compared to BT. Hence the tissue model with artery and vein responds to external pulsed heating more stiffly with higher strain rate compared to the bare tissue model as displayed in Fig. 9. In contrast the stress-strain curve of continuous mode heating suggests an equally stiffed tissue medium for both the CVTT and BT model (Fig. 9).

The faster thermal and strain response of tissue under continuous heating induces more internal thermal stress compared to pulsed heating (Fig. 10 [Column 2, 3]). Therefore, it suggest that a more thermal pain is evolved during the shorter continuous heating with respect to longer pulsed heating for a limited peak temperature of tissue. Also the improved acoustic and optical characteristic of tissue with nanoparticle doping caused more thermally induced internal stress under external heating (Fig. 10). The presence of arteries and veins in the tissue model affects the induced stress profile similar to the temperature profile (Fig. 10, 6) qualitatively. The reduced internal stress induced under the same thermal load for unfocused laser source suggests a lower thermal pain with respect to focused ultrasound source. The tissue deformed more under pulsed heating due to wider therapeutic temperature history compared to continuous agitation (Fig.6, 11). The dispersed therapeutic heat of unfocused energy source causes a wider deformation compared to focused energy source. Hence the tissue medium becomes less stiffy during the unfocused heating in comparison to the focused heating. In the present study, the tissue medium tends to thermally deform in upward direction (Fig. 11) as the bottom surface of the computational domain is set fixed.

## VI. CONCLUSIONS

In this study, the thermal and mechanical responses of tissue have been compared between the continuous and pulse mode heating during HIFU and LITT. The goal is to obtain a precise tumor necrosis with lower thermally induced nociceptive pain. The non-Fourier model shows significant changes in the thermal and damage responses of tissues during focused ultrasound heating especially for shorter period of heating. With limited rise in the peak tissue temperature, the pulsed mode heating for longer period with lower duty cycle (16.6%) shows a wider span of threshold thermal energy compared to the continuous mode heating. Indeed the tissue temperature rise in hysteretic manner results a target specific necrotic damage distribution preserving the surrounding healthy tissue in contrast to continuous mode of heating. The non-homogeneous effect of multilevel artery and vein in tissue has been found to be prominent during LITT compared to HIFU. Most importantly, the effect of artery-vein has been found to be significant during pulsed mode of heating in contrast to the continuous mode. Besides a reduced thermoelastic stiffness of tissue is revealed under pulse heating due to lower thermally induced strain rate in respective of continuous heating. As a result, the thermally induced nociceptive pain in tissue have been found to be more for continuous agitation of energy compared to pulse mode. Additionally, the presence of large blood vessels elevates the stiffness of tissue to react to external thermal load. The enhanced optical and acoustic characteristics of tissue in presence of nanoparticles affect both the thermal and elastic behavior of tissue under external heating. Moreover, the dispersed energy source results a wider thermally induced tissue deformation in contrary to focused energy source. Nevertheless, the above findings can help to achieve effective thermal ablation of tumor during pre-clinical trial.

ACKNOWLEDGMENTS

We acknowledge Dr. Himanshu Shekhar for providing his lab facilities at IIT Gandhinagar. We also acknowledge the guidance of Dr. Himanshu Shekhar and Dr. Dipesh Shah for present experimental work.

**Abhijit Paul**

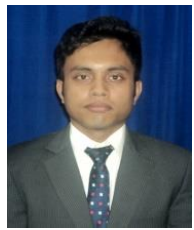

Abhijit Paul is pursuing his Ph.D. degree from NIT Arunachal Pradesh and also a SRF in SERB (DST) sponsored project, since May 2019. He has completed his M.Tech degree from NIT Agartala in 2016 with 9.33 CGPA. He remained gold medallist of B.E batch 2014, Mechanical Engineering Department, Tripura University. He has qualified GATE in 2014.

**Dr. Anup Paul**

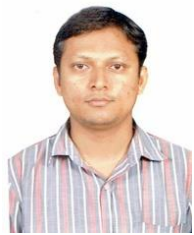

Dr. Anup Paul is an Assistant Professor in Mechanical Engineering Department at National Institute of Technology Arunachal Pradesh, India since Dec. 2015. He has received Ph.D. from Indian Institute of Technology Madras, (IIT Madras) in 2015, Master in Mechanical Engineering in 2008 and Bachelors in Mechanical Engineering in 2005. Prior to joining NIT Arunachal Pradesh, he was working as an Assistant Professor at Tezpur University, India.